\documentclass{aa}

\usepackage{graphicx}
\usepackage{txfonts}
\usepackage{natbib}

\begin{document}

\title{Characterising stellar micro-variability for planetary transit
  searches}

\author{S. Aigrain\inst{1} \and F. Favata\inst{2} \and G.
  Gilmore\inst{1} }

\offprints{S. Aigrain}

\institute{Institute of Astronomy (IoA), University of Cambridge,
  Madingley Road, Cambridge CB3 0HA, United Kingdom \\
  \email{suz@ast.cam.ac.uk, gil@ast.cam.ac.uk} \and Astrophysics
  Division, Research \& Scientific Support Department,
  European Space Agency, P.O. Box 299, 2200 AG Noordwijk, The Netherlands\\
  \email{Fabio.Favata@rssd.esa.int} }

\date{Received \ldots; accepted \ldots}

\abstract{ A method for simulating light curves containing stellar
  micro-variability for a range of spectral types and ages is
  presented. It is based on parameter-by-parameter scaling of a
  multi-component fit to the solar irradiance power spectrum (based on
  VIRGO/PMO6 data), and scaling laws derived from ground based
  observations of various stellar samples.
     
  A correlation is observed in the Sun between the amplitude of the
  power spectrum on long (weeks) timescales and the BBSO Ca\,{\sc ii}
  K-line index of chromospheric activity. On the basis of this
  evidence, the chromospheric activity level, predicted from rotation
  period and $B-V$ colour estimates according to the relationship
  first introduced by \citet{noy83} and \citet{nhb+84}, is used to predict the
  variability power on weeks time scale. The rotation period is
  estimated on the basis of a fit to the distribution of rotation
  period versus $B-V$ observed in the Hyades and the \citet{sku72}
  spin-down law. The characteristic timescale of the variability is
  also scaled according to the rotation period.
  
  This model is used to estimate the impact of the target star
  spectral type and age on the detection capability of space based
  transit searches such as \emph{Eddington} and \emph{Kepler}. K stars
  are found to be the most promising targets, while the performance
  drops significantly for stars earlier than G and younger than 2.0
  Gyr. Simulations also show that \emph{Eddington} should detect
  terrestrial planets orbiting solar-age stars in most of the
  habitable zone for G2 types and all of it for K0 and K5 types.
  
   }

\maketitle


\section{Introduction}
\label{sec:intro}

The transit method is reaching maturity as a method to discover
gaseous giant exo-planets from the ground. Several transit-search
projects have now produced convincing candidates, which are awaiting
confirmation through other methods.

However, the few years of experience now available in this field have
exemplified the difficulty of the task at hand. The signal is small
($\approx 2$ \% at most), short (a few hours), with periods ranging
from a few days to several years, and is embedded in noisy data
(photon, sky, background, instrumental noise, etc\ldots). Ground-based
projects also suffer from the daily interruptions in the observations
and the finite duration of the observing runs, and are affected by
atmospheric scintillation noise. A variety of transit search
algorithms have been developed, ranging from a simple matched filter
to more sophisticated methods \citep{ddk+2000,ddb2001,af02,kzm02}. All
of these are optimised to work in the presence of white noise, and
tests of simulated data have shown they can perform very well,
reliably detecting transits at the extreme margin of statistical
significance. Rather than the detection of the transits themselves,
the major difficulty for ground-based searches has in fact been
distinguishing transit-like events caused by stellar systems, such as
eclipsing binaries with high mass ratios, or hierarchical triple
systems (due to either a physical triple system or and eclipsing
binary blended with a foreground star), from true planetary transits.

\begin{figure*}
  \centering \includegraphics[width=\textwidth]{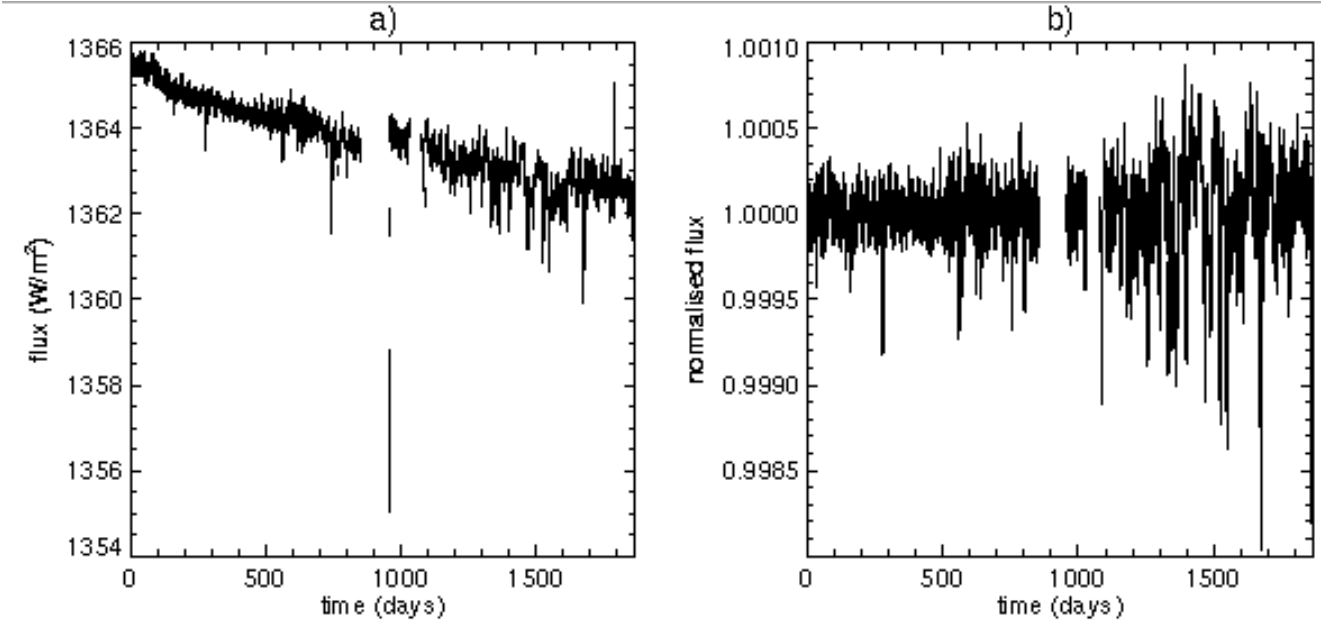}
  \caption{PMO6 light curve, {\bf a)} before and {\bf b)} after the 
    pre-processing steps described in Sect.\ \ref{sec:dataproc}. The
    data starts in January 1996. \label{fig:dpmo}}
\end{figure*}

In order to detect terrestrial planets, it is necessary to go to
space, to avoid being affected by atmospheric scintillation and to
monitor the target field(s) continuously, with minimal
interruptions. However, with improved photometric precision comes an
additional noise source, which is usually insignificant at the
precisions achieved by ground based observations: the intrinsic
variability of the stars, due mainly to the temporal evolution and
rotational modulation of structures on the stellar disk. The Sun's
total irradiance (see Fig.\ \ref{fig:dpmo}) varies on all timescales
covered by the available data, with a complex, non white power
spectrum (see Fig.\ \ref{fig:rpmo}). The amplitude of the variations
can reach more than $1$ \% when a large spot crosses the solar disk at
activity maximum, compared to transit depths of tenths to hundredths
of a percent. There is significant power on timescales of a few hours,
similar to the typical transit duration. Thus Sun-like variability, if
left untreated, would significantly affect the detection performance
of missions such as \emph{Eddington} or \emph{Kepler} \citep{agf01}. 

\emph{Eddington} \citep{fav03} is an ESA mission to be
implemented as part of the Agency's \ ``Cosmic Vision'' science
program, with a planned launch date in 2007/2008. Its two primary
goals are the study of stellar structure and evolution through
asteroseismology, and the detection of habitable planets via the
transit method. \emph{Kepler} \citep{bkl+03} is a NASA Discovery
mission which will concentrate primarily on the second of these two
goals, and is planned for launch in 2008.

A number of algorithms designed to reduce the impact of
micro-variability have been tested on simulated data
\citep{ddb2001,jen02,caf03}.  Some are modified detection algorithms,
designed to differentiate between the transit signal and the stellar
noise. Others are pre-processing tools, which whiten the noise profile
before running standard transit detection algorithms. Inserting solar
irradiance data from VIRGO into simulated light curves, \citet{caf03}
showed that the transit detection performance achieved in the presence
of Sun-like variability after pre-processing can be as good as that
obtained with unprocessed light curves containing white noise only.
\citet{jen02} applied a simple scaling to the solar irradiance data to
evaluate the impact of increased rotation rate.  Nonetheless, a more
physical model, in which the different phenomena involved can be
scaled independently in timescale and amplitude for a range of
spectral types and ages, is needed to simulate realistic light curves
for stars other than the Sun. This will allow us to optimise, evaluate
and compare different algorithms, but also different design and target
field options for the space missions concerned.

The present paper is concerned with the development of such a model.
The philosophy adopted in the process is the following. Intrinsic
stellar variability is by no means a well-understood process. Despite
recent progress in the modelling of activity-induced irradiance
variations on timescales of days to weeks in the Sun
\citep{ksf+03,lrp+03}, the extension of these physical models to other
stars remains problematic, due to the scarcity of information on how
the timescales, filling factors of various surface structures, and
contrast ratios, depend on stellar parameters. We have therefore
adopted an empirical approach, using chromospheric flux measurements
as a proxy measure of activity-induced variability. This step is
possible due to the fact that a correlation between the two quantities
is observed in the Sun throughout its activity cycle, as well as in
other stars. Similarly, empirically derived relationships were used
again to relate chromospheric activity, rotation, age and colour,
rather than attempting to use models which make a number of
assumptions about the physical process driving these phenomena, and
generally depend on parameters which require fine-tuning.

The analysis of the evolution of the Sun's total irradiance variations
with the solar activity cycle is described in Sect.\ \ref{sec:sun}:
the power spectrum of the variations is modelled as a sum of
powerlaws, whose parameters are tracked as they evolve throughout the
solar cycle. The same type of model is used to construct artificial
power spectra and light curves for stars with given theoretical
parameters (age, spectral type). A number of empirically derived
scaling laws are used to relate these input stellar parameters to the
power spectrum parameters (Sect.\ \ref{sec:model}). As a consistency
check and an illustration of the possible applications of this model,
the results of a small set of simulations designed to evaluate the
impact of variability from different stars on \emph{Eddington}'s
planet-hunting capabilities are described in Sect.\ \ref{sec:simul}.


\section{Clues from solar irradiance variations}
\label{sec:sun}

The Sun is the only star observed with sufficient precision and
frequent sampling to permit detailed micro-variability studies, thanks
to the recent wealth of data collected by the SoHO spacecraft, and
particularly the full disk observations obtained by VIRGO (Variability
of solar IRradiance and Gravity Oscillations), the experiment for
helioseismology and solar irradiance monitoring on SoHO,
\citep{faa+97}.

Stellar micro-variability is difficult to observe from the ground due
to its very low amplitude, except for very young, active stars --
which are outside the main range of interest for planet searches.
There is some information available on rms night-to-night and
year-to-year photometric variability of a small sample of stars
monitored over many years by a few teams \citep{rls+98,hbd+00}. We
make use of these as they present the advantage of covering a range of
stellar ages, but their irregular time coverage and limited
photometric precision make them unsuitable for an in-depth study, and
particularly for the detailed analysis of the frequency content of the
variations.

A drastic improvement in our understanding of intrinsic stellar
variability across the HR diagram is expected from the very missions
this work is aimed at preparing. In the relatively short term, MOST
will provide valuable information for a small sample of stars, but it
is not until the launch of COROT, and later \emph{Kepler} and
\emph{Eddington}, that a wide range of stellar parameters will be
covered. In the mean time, we must make use of the detailed solar
data, and make reasonable assumptions to extrapolate to other stars
than the Sun.

\subsection{SoHO/VIRGO total irradiance (PMO6) data}
\label{sec:data}

All SoHO/VIRGO data used in this work were kindly provided by the
VIRGO team. The main instrument of interest was PMO6, a
radiometer measuring total solar irradiance.

The light-curves cover the period January 1996 to March
2001\footnote{Except for two interruptions roughly 1,000 days after
the start of operations, corresponding to the ``SoHO vacations'',
when the satellite was lost and then recovered.}, which roughly
corresponds to the rising phase of cycle 23.

\subsubsection{Pre-processing of the data}
\label{sec:dataproc}

The light curves were received as level 1 data, in physical units but
with no correction for instrumental effects {\bf or outliers}. Careful
treatment was required to remove long term (instrument decay) trends.
There was a difference of $\sim 0.24$\,\% in the mean measured flux
between the start and the end of the time series. Given that the
observations roughly correspond to the interval between the minimum
and the maximum of the Sun's activity cycle, one might expect to see a
rise in the mean irradiance over that period. The instrumental decay
may therefore be higher than the value quoted. However, the absolute
value of the irradiance was of little interest for the present study,
which concentrates on relative variations on time scales of weeks or
less. Any long term trends in the data were therefore removed
completely, regardless of whether they were of instrumental or
physical origin. The decay appeared non-linear and there were
discontinuities and outliers in the light curves, making a simple
spline fit unsuitable.

The approach that was adopted consisted of a 3 step process:
\begin{itemize} 
\item Obvious outliers were removed by computing residuals from a
  spline fit (whose nodes were spaced two months apart to avoid being
  sensitive to the Sun's rotation period) and applying a 1 $\sigma$
  cutoff.
\item Visual inspection of the data then allowed us to manually remove
  sections visibly affected by instrumental problems.
\item Spline fits were performed on intervals chosen by visual
  inspection to start and end where discontinuities occurred. Each
  interval was divided by the corresponding fit, resulting in a
  normalised output light curve.
\item A 5 $\sigma$ cutoff was applied for outlier removal.
\item The sampling, originally 1 min, was reduced to 15 min to make
  the size of the light curves more manageable. This was done by
  taking the mean of the original data points in each 15~min bin,
  ignoring any missing or bad data points. It is unlikely any
  information on timescales shorter than 15 min would significantly
  impact the transit detection process, as the transits of interest
  here generally last several hours (corresponding to orbital periods
  of several months or years).
\item Data gaps were replaced with the baseline value of $1.0$, to
  allow the calculation of the power spectra needed for the analysis.
\end{itemize}

\subsection{Modelling the `solar background'}
\label{sec:sbm}

The power spectrum of the solar irradiance variations at frequencies
lower than $\simeq 8$~mHz constitutes a noise source for
helioseismology, usually referred to as the `solar background'. It is
common practice to fit this background with a sum of powerlaws in
order to model it accurately enough to allow the measurement of solar
oscillation frequencies and amplitudes. Powerlaw models were first
introduced by \citet{har85}. The most commonly used model in the
literature today is that of \citet{alt94}, which is fairly similar:
the total power spectrum is approximated by a sum of power laws, the
number $N$ of which varies between three and five depending on the
frequency coverage:
\begin{equation}
 \label{eq:powlaw}
      P(\nu) = \displaystyle{\sum_{i=1}^{N}} P_{i} = 
                \displaystyle{\sum_{i=1}^{N}} \frac{A_i}
                {1+(B_i~\nu)^{C_i}}
\end{equation}
\noindent where $\nu$ is frequency, $A_i$ is the amplitude of the
$i^{\mathrm{th}}$ component, $B_i$ is its characteristic timescale,
and $C_i$ is the slope of the power law (which was fixed to $2$ in
Harvey's early model). For a given component, the power remains
approximately constant on timescales larger than $B$, and drops off
for shorter timescales.  Each power law corresponds to a separate
class of physical phenomena, occurring on a different characteristic
time scale, and corresponding to different physical structures on the
surface of the Sun (see Table \ref{tab:comps}).

\begin{table} 
  \caption[]{Typical timescales for the different of structures on the 
    solar surface. \label{tab:comps}}
      $$
         \begin{array}{p{0.5\linewidth}l}
            \hline
            \noalign{\smallskip}
            Component & \mathrm{Timescale}~B~(\mathrm{s}) \\
            \noalign{\smallskip}
            \hline
            \noalign{\smallskip}
            Active~regions     & 1~\mathrm{to}~3 \times 10^7 \\
            Super-granulation  & 3~\mathrm{to}~7 \times 10^4 \\
            Meso-granulation   & \simeq 8000 \\
            Granulation        & 200~\mathrm{to}~500 \\
            Bright~points      & \simeq 70 \\
            \noalign{\smallskip}
            \hline
         \end{array}
      $$
\end{table}

\subsection{Evolution of the power spectrum with the activity cycle}
\label{sec:sbox}

\begin{figure*}
  \centering \includegraphics[width=\textwidth]{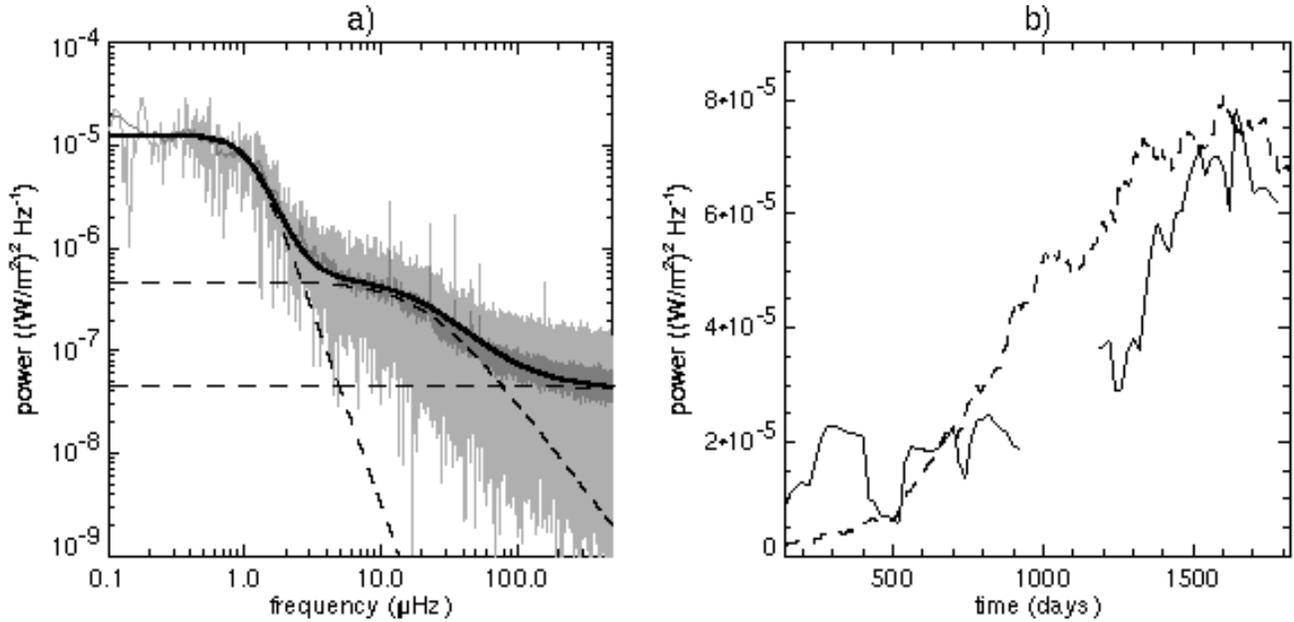}
  \caption{{\bf a)} Power spectrum of the PMO6 light curve (1996-2001). 
    Light grey: power spectrum. Dark grey: idem, smoothed with a
    boxcar algorithm. Thick solid line: multi-component powerlaw fit
    (see Sect. \ref{sec:sbm}). Dotted lines: individual components of
    the fit. \hspace{0.0cm} {\bf b)} Comparison between the the time
    dependence of the amplitude of the low frequency component of the
    power spectrum ($A_1$) and chromospheric activity.  Solid line:
    Evolution of $A_1$, computed as described in Sect.\ 
    \ref{sec:sbox}, using $L = 180$ days and $S = 20$ days, between
    1996 and 2001. The gap at around 1000 days corresponds to a
    prolonged gap in the data.  Dotted line: BBSO Ca\,{\sc ii} K-line
    index over the same period (arbitrary units), smoothed with a
    boxcar algorithm (base $180$~days).
    \label{fig:rpmo}}
\end{figure*}

In order to track the evolution of the solar background over the
activity cycle, sums of power-laws -- as given by Eq.\ \ref{eq:powlaw}
-- were fitted to the power spectrum of a section of data of duration
$L$ (typically 6 months).  Such a fit is illustrated on the power
spectrum of the entire dataset in the left-hand panel of Fig.\ 
\ref{fig:rpmo}.  The operation is then repeated for a section
shifted by a small interval $S$ from the previous one (typically 20
days), and so on. Thus the evolution of each component can be tracked
throughout the rise from solar minimum (1996) to maximum (2001) by
measuring changes in the parameters defining each powerlaw.

A single component fit with parameters $A_1$, $B_1$ \& $C_1$ is made
first. Additional components are then added until they no longer
improve the fit, i.e.\ until the addition of an extra component does
not reduce the $\chi^2$ by more than $10^{-2}$. The fit to the first
section is used as the initial guess for the fit to the next section,
and so forth. This method allows us to track the emergence of
components corresponding to different types of surface structures
throughout the solar cycle, as well as monitor variations in
amplitude, timescale and slope for each component.

\begin{figure}
  \centering \includegraphics[width=\linewidth]{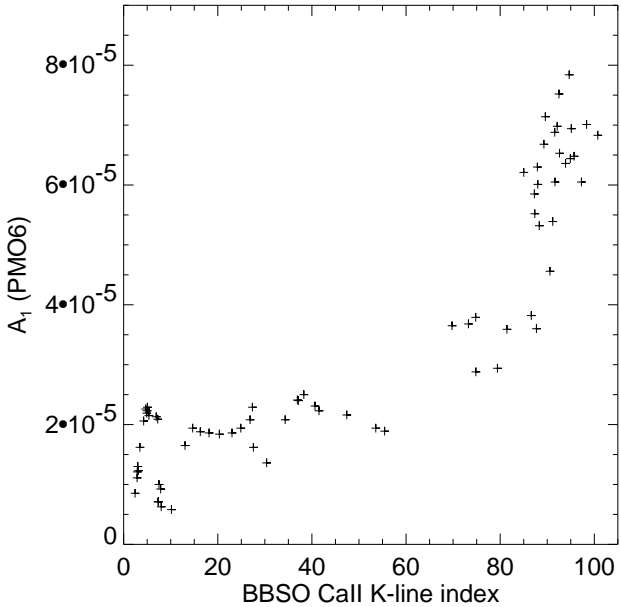}
  \caption{Scatter plot of the amplitude of the low frequency component of 
    solar irradiance variations $A_1$ (computed as described in Sect.\ 
    \ref{sec:sbox}, using $L = 180$ days and $S = 20$ days), versus
    the BBSO Ca\,{\sc ii} K-line index (arbitrary units) over the
    period 1996 to 2001.
  \label{fig:rpmo_corr}}
\end{figure}

\subsubsection{Results} 
\label{sec:rvirgo}

The algorithm described above was run on the PMO6 data with $L = 180$
days and $S = 20$ days, and three components were found to provide the
best fit in all cases. These components have stable timescales and
slope, varying in amplitude only. The physical processes giving rise
to each component are thus of a permanent nature.  A number of points
of interest emerge from the results.  The first component, with $\tau
\simeq 1.3 \times 10^{5}$~s (active regions) shows an increasing trend
in amplitude which is well correlated with the Ca\,{\sc ii} K-line
index, an indicator of chromospheric activity. This is illustrated in the
right-hand panel of Fig.\ \ref{fig:rpmo}. The slope of the powerlaw
is 3.8 (in good agreement with \citealp{aac98}).

The observed correlation, which is further illustrated by the scatter
plot in Fig.\ \ref{fig:rpmo_corr} comes as no surprise. The passage of
individual active regions across the disks of the Sun and other stars
monitored by the Mt Wilson HK Project can be clearly seen in plots of
the activity index $S$ (from which $R'_{\mathrm{HK}}$ is derived)
versus time\footnote{See the Mt Wilson HK project homepage, {\tt
    http://www.mtwilson.edu/Science/HK\_Project/}.}.  On the other
hand, the effect of the same type of event on the solar irradiance has
been studied with a number of instruments, most recently VIRGO/LOI and
PMO6 \citep{dsa+98}. Recent models including contributions from
faculae and sunspots of tunable size and number reproduce the PMO6
light curve to a high degree of precision \citep{ksf+03,lrp+03}.

However, observing and characterising a correlation throughout the
Sun's activity cycle, between a chromospheric activity indicator which
can be measured from the ground for a large number of stars, and total
irradiance variations, whose amplitudes are so small they are only
observable by dedicated high-precision photometric space missions,
goes one step further. Most importantly for planetary transit
searches, it implies that chromospheric activity indicators such as
$R'_{\mathrm{HK}}$ can be used as a \emph{proxy} to predict weeks
timescale variability levels for a wide range of stars.

The amplitude of the second component also increases, but is not
correlated to the Ca\,{\sc ii} index. This component corresponds to
timescales corresponding to super- and meso- granulation, for which no
detailed models are available to date. Our understanding of this kind
of phenomenon is expected to improve dramatically when the results
from space-based experiments designed for precision time-series
photometry will become available.

\subsubsection{Implications}
\label{sec:ivirgo} 

The correlation between $A_1$ and the Ca\,{\sc ii} K-line index,
although not extremely tight, is a clear indication that chromospheric
activity indicators contain information about the variability level of
the Sun on timescales longer than a few days. To establish more
solidly a scaling law between photometric variability and
chromospheric activity, we must use a wider stellar sample to
constrain the relation over the entire expected range of activity
levels (see Sect.\ \ref{sec:rms2y}).

Little useful information has been extracted from the solar data on
what determines the parameters of the solar background other than
$A_1$. The few clues available from other sources will be presented in
Sect.\ \ref{sec:other}.


\section{Modelling stellar micro-variability}
\label{sec:model}

The main aim of the present model is the simulation of realistic light
curves of stars more active than the Sun, if possible as a function of
stellar parameters such as spectral type and age, via observables such
as $B-V$ colour and rotational period $P_{\mathrm{rot}}$.  The
multi-component power-law model used in Sect.\ \ref{sec:sun} to fit
the solar background power spectrum now forms the basis of the
simulation of enhanced variability light curves.
   
To simulate a light curve for a given star, the first task is to
generate a power spectrum using this power-law model. When applying the
component-by-component procedure described in Sect.\ \ref{sec:sbox} to
fit solar power spectra, the optimal number of components was found to
be three. We therefore use a sum of three power-laws to generate the
stellar power spectra. The highest frequency component, a
superposition of granulation, oscillations and white noise, has a
characteristic timescale which is shorter than the typical sampling
time for planetary transit searches, and thus can be replaced by a
constant value. There are therefore a total of 7 parameters to adjust
for each simulated power spectrum: three for each resolved power-law
plus one constant.

The inputs of the model are the spectral type and age of the star.
Starting from these theoretical quantities given, how are the
power-law parameters deduced?  Most of the information available
concerns the amplitude of the first power-law, $A_1$. We have
established in Sect.\ \ref{sec:sun} that there is a correlation
between $A_1$ and the Ca~{\sc II} K-line indicator of chromospheric
activity. On the other hand, there is a well known scaling between
rotation period, colour and chromospheric activity
\citep{nhb+84}. Provided one can estimate the rotation period of the
star (see Sect.\ \ref{sec:ptbv}), the activity level is computed
(Sect.\ \ref{sec:rpbv}), and from this one obtains $A_1$ (Sect.\
\ref{sec:rms2y}).  Sect.\ \ref{sec:other} summarises how the other
parameters of the power-law model, on which much less information is
available, are estimated.
 
\subsection{The rotation period-colour-age relation}
\label{sec:ptbv}

As detailed in Sect.~\ref{sec:rpbv}, it is possible to deduce the
expected activity level for a star of known mass (i.e.\ colour) and
rotation period.  However, the number of stars with known rotation
periods is relatively small. In the context of the present work, it
would thus be useful to be able to predict the rotation period for a
given stellar mass and age.

Observational constraints on rotation rates for stars of known mass
and age come from two sources: star forming regions and young open
clusters, where one can measure photometric rotation periods or
rotational line broadening ($v \sin i$); and the Sun itself. There is
little else, as rotational measurements are hard to perform for all
the quiet, slowly rotating intermediate age and old stars other than
the Sun, except for relatively nearby field stars, for which little
reliable age information is available. The status of observational
evidence and theoretical modelling in this domain is outlined by
\citet{kpb+97}, \citet{bfa97} and \citet{st03}. Here we briefly sketch
the current paradigm to set the context of the present work.

The observed initial spread in rotation velocities (from measurements of
T-Tauri stars, with a concentration around $10$--$30$~km\,s$^{-1}$ but
a number of fast rotators, up to $100$'s of km\,s$^{-1}$) is
attributed to the competing effects of spin-up (due to the star's
contraction and accretion of angular momentum from the disk) and
slowing-down mechanisms such as disk-locking \citep{koe91,bfa97}. This
spread is observed to diminish with age (by the age of the Hyades, only
some M-dwarfs still exhibit fast rotation, \citealt{psd+95}), leading to
a dependency of rotation on mass only. Following this homogenisation,
one observes constant spin-down in a given mass range (cf.\ the
\citealt{sku72} $t^{1/2}$ law for Sun-like stars). Both homogeneisation
and power-law spin-down can be explained by the loss of angular
momentum through a magnetised wind \citep{sch62,wd67}, a mechanism
which is more effective in faster rotators.

Angular momentum evolution models have vastly improved recently, but
they still rely on the careful tuning of a number of parameters,
especially for young and low-mass stars. We have therefore chosen to
use empirically derived scaling laws and to restrict ourselves to the
range of ages (older than the Hyades) and spectral types (mid-F to
mid-K) where a unique colour-age-rotation relation can be
established. In this range the forementioned parameters become less
relevant and the models reproduce the observations fairly robustly.

A relationship between $B-V$ colour (i.e.\ mass) and rotation at a
given age was empirically derived from photometric rotation period
measurements in the Hyades \citep{rtl+87, rls+95}.  Only rotation
periods were used, rather than $v \sin i$ measurements, to avoid
introducing the extra uncertainty of assigning random inclinations to
the stars and having to assume theoretical radii to convert $v \sin i$
to a period. This relationship is valid for the range $0.45 \leq B-V
\leq 1.3$. For stars bluer than $B-V = 0.45$, rotation rates saturate,
but this is outside the range of spectral types of interest for the
present work. Redder than $B-V = 1.3$, a significant spread is still
observed in the rotation periods. The remaining range is divided into
two zones, each following a linear trend. Redder than $B-V = 0.62$,
the slope of the relation is quite close to the theoretical relation
obtained by \citet{kaw89}. We also estimated the age of the Hyades
from the zero-point of the linear fit as done by \citet{kaw89}, but
incorporating the improved data from \citet{rls+95}. The age obtained
in this manner is $634$~Myr, consistent with recent determinations by
independent methods: $655$~Myr \citep{cay90}, $600$~Myr \citep{tsl97},
and $625$~Myr \citep{pbl+98}, thereby confirming the quality of the
fit. The slope for stars bluer than $B-V = 0.62$ is much steeper,
presumably due to the thinner convective envelopes of the stars in
this range.

\begin{figure}
  \centering \includegraphics[width=\linewidth]{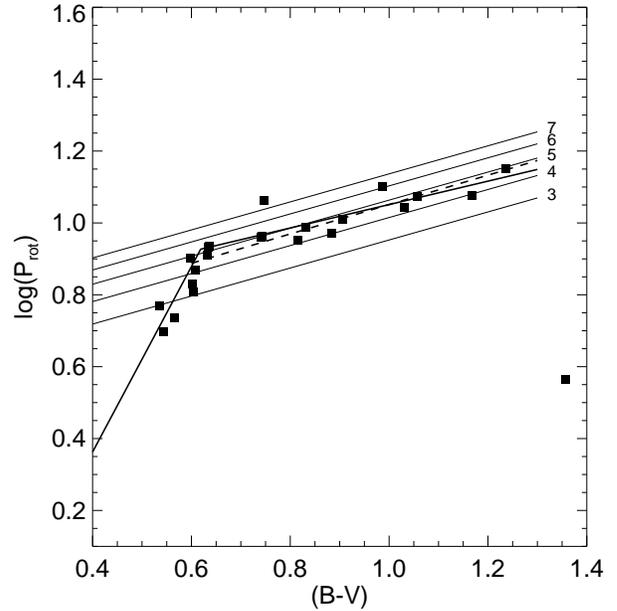}
  \caption{Plot of rotation period versus $B-V$ colour for Hyades
    stars. Data from \citet{rtl+87,rls+95}. The thin parallel black
    lines correspond to `rotational isochrones' from \citet{kaw89}
    with $t$ as indicated next to each line (in units of $10^8$~yr).
    The dashed line is a linear fit to the data with $0.6 \leq B-V <
    1.3$. The thick solid line is a composite of two linear fits, one
    for $B-V < 0.62$ and one for $0.62 \leq B-V < 1.3$.
    \label{fig:p2bv}}
\end{figure}

This can then be combined with the $t^{1/2}$ spin-down law into a
rotation-colour-age relation:
\begin{equation}
 \label{eq:bvt2p}
      \begin{array}{l}
        \log \left( P_{\mathrm{rot}} \right) 
        - 0.5~\log \left( \frac{t}{625~\mathrm{Myr}} \right) =  
        \vspace{0.3cm} \\
        \hspace{0.5cm} \left\{ 
        \begin{array}{rr}
          - 0.669 + 2.580~(B-V), \hspace{0.5cm} 
        & 
          0.45 \leq B-V < 0.62 \\
          0.725 + 0.326~(B-V), \hspace{0.5cm} 
        & 
          0.62 \leq B-V < 1.30
        \end{array} 
        \right\}
      \end{array}
\end{equation}
   
A comment on the adopted value of $0.5$ for $n_t$, the index in the
spin-down law is appropriate. It has recently been suggested that a
$0.6$ might be more appropriate (\citealt{gr02}, on the basis of an
updated sample of Sun-like stars with some new age
determinations). However, the original value of $0.5$ was kept for the
present work.  The change would not affect the predicted rotation
rates significantly, and the errors on this new value of $n_t$ (which
depends, for example, on age determinations from isochrone fitting)
are larger than the difference. We have therefore kept the lower
value, as it leads, if in error, to overestimated rotation rates,
hence more variability and on faster timescales, and eventually
conservative estimates of transit detection rates.

\subsection{The activity-rotation period-colour relation}
\label{sec:rpbv}

The next step consists in estimating from the colour and rotation
period the expected chromospheric activity level of the star. For this
purpose, the scaling law first derived by \citet{noy83} and
\citet{nhb+84} is used. It relates the Ca\,{\sc ii} index $\left<
R'_{\mathrm{HK}} \right>$ to the inverse of the Rossby number
$R_{\mathrm{o}}$, and can thus be understood in terms of the interplay
between convection, rotation and the star's dynamo: 
\begin{equation}
 \label{eq:rhk2ro}
      - \log R_{\mathrm{o}} = 0.324 - 0.400~y + 0.283~y^2 - 1.325~y^3
\end{equation}
\noindent where $y = \log \left< R'_{\mathrm{HK}} \right>_5 $ and
$\left< R'_{\mathrm{HK}} \right>_5 = \left< R'_{\mathrm{HK}} \right>
\times 10^5$.  $R_{\mathrm{o}}$ is related to the rotation period
$P_{\mathrm{rot}}$ and $B-V$ colour as follows:
\begin{equation}
 \label{eq:ro2ptau}
      R_{\mathrm{o}} = \tau_{\mathrm{c}} / P_{\mathrm{rot}}
\end{equation}
\noindent where $P_{\mathrm{rot}}$ is expressed in days, and the
following (empirically derived) relation for $\tau_{\mathrm{c}}$, the
convective overturn time, is used:
\begin{equation}
 \label{eq:tauc}
      \begin{array}{l}
        \log \left( \tau_{\mathrm{c}} \right) = 
        \vspace{0.3cm} \\
        \hspace{0.5cm} \left\{
        \begin{array}{lcl}
          1.361 - 0.166~x + 0.025~x^2 - 5.323~x^3, & & x \geq 0 \\
          1.361 - 0.140~x,                         & & x < 0
        \end{array} 
        \right\}
      \end{array}
\end{equation}
\noindent where $x = 1 - (B-V)$. Equation (\ref{eq:rhk2ro}) was
inverted (using an interpolation between tabulated values) to allow us
to deduce the chromospheric activity index from the rotation period
and $B-V$ colour of a star. For details of how the above relations,
which are simply stated here, were obtained, the reader is referred to
\citet{nhb+84}.

\subsection{Active regions variability and chromospheric activity}
\label{sec:rms2y}

Having noted that the `active regions' component of the solar
activity spectrum appears directly correlated to chromospheric activity (see
Sect.\ \ref{sec:sbox}), we turn to the small but valuable datasets
containing both photometric variability measurements and activity
indexes for a variety of stars. The data available on any one star is
of course much less precise and less reliable than the solar data, but
the dataset as a whole spans a much wider range of activity levels.

We use a sample of stars for which both $\left< R'_{\mathrm{HK}}
\right>$ and rms photometric variability measurements are available in
the litterature.  This sample, mainly taken from \citet{rls+98} and
\citet{hbd+00}, covers a wide range of spectral types and ages,
including the active young (130 Myr) solar analogue EK Dra and some
older, Sun-like stars which are known to harbour planets from radial
velocity measurements.

These data can be used to derive a quantitative relationship between
$\left< R'_{\mathrm{HK}} \right>$ and night-to-night rms variability
(in Str\"omgren $b$ and $y$; $\mathrm{rms}_{by} =
\mathrm{rms}\{(b+y)/2\}$).  All rms values used are in units of
$10^{-4}$~mag.
  
\begin{figure}
  \centering \includegraphics[width=\linewidth]{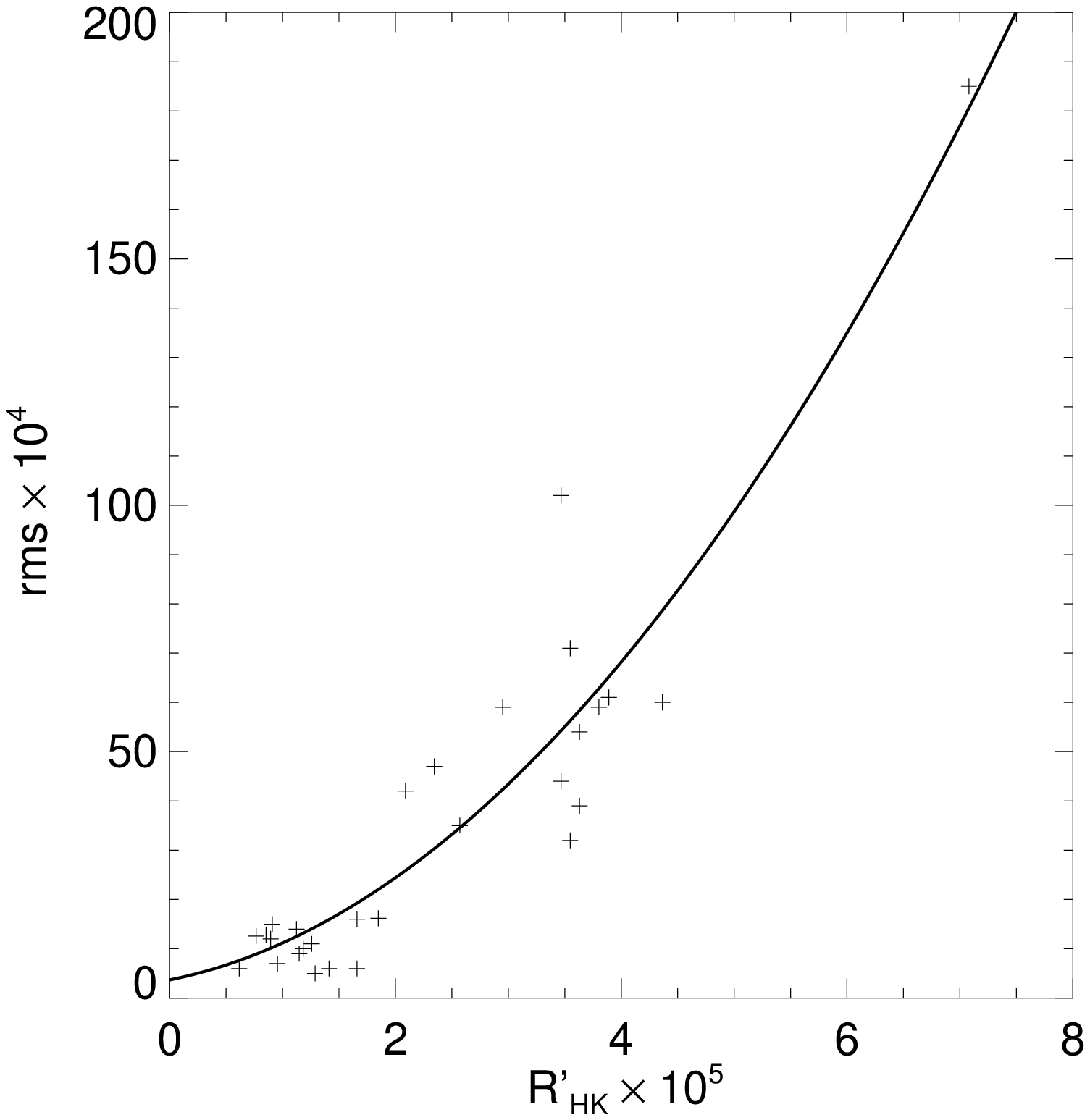}
  \caption{Plot of $\mathrm{rms}_{by}$ versus $\left<
      R'_{\mathrm{HK}} \right>_5$ for a sample of stars from
    \citet{rls+98} and \citet{hbd+00}, with $2^{\mathrm{nd}}$ order
    polynomial fit overlaid. \label{fig:rms2y}}
\end{figure}

As illustrated in Fig.\ \ref{fig:rms2y}, a $2^{\mathrm{nd}}$ order
polynomial provides a good fit to the relationship between
$\mathrm{rms}_{by}$ and $\left< R'_{\mathrm{HK}} \right>_5$. The
equation of the fit is:
\begin{equation}
 \label{eq:rms2y}
     \mathrm{rms}_{by} = 
       3.69 + 4.60~\left< R'_{\mathrm{HK}} \right>_5
         + 2.88~\left< R'_{\mathrm{HK}} \right>_5^2
\end{equation}

Note that one star, EK Dra, is significantly more active and variable
than the rest of the sample -- because it is younger. Including it in
the fit thus gives it a disproportionate weight, but we have included
it because it broadens the range of $\left< R'_{\mathrm{HK}} \right>$
covered by a factor of 2. Furthermore, previous studies have shown
\citep{mg02} that it fits tightly on relationships between activity
and stellar parameters derived from samples of solar analogues of
various ages, suggesting its behaviour is representative of the
mechanisms driving activity in general -- and thus, in our reasoning,
variability on the timescales under consideration.

To be usable for our purposes, Eq.~(\ref{eq:rms2y}) must be completed
by relationship between $\mathrm{rms}_{by}$ and $A_1$. The desired
value of $A_1$ corresponds to white light relative flux variations,
not $b$ and $y$ magnitude variations. To obtain the rms of the
relative flux variations one must multiply the rms of the magnitude
variations by a constant factor of $2.5 / \ln(10) = 1.08$. This must
then be converted to a white light flux rms value. This requires the
definition of a reference point in the solar cycle, at which to
compare the variability levels in the two bandpasses. \citet{rls+98}
quote a single $\log \left< R'_{\mathrm{HK}} \right>$ value of $-4.89$
for the Sun, which is an average of measurements performed over many
years. This defines a reference solar activity level, corresponding
roughly to a third of the way into the rising phase of cycle 23. The
average night-to-night variability as measured in $b$ and $y$ by
\citet{rls+98} is $\mathrm{rms}_{by} = 5 \times 10^{-4}$~mag. The
corresponding white light variability level, measured from a 6 month
long section of PMO6 data downgraded to 1 day sampling, centred on the
date for which the measured activity level was equal to the reference
level defined above, is $\mathrm{rms_{white}} \approx 1.8 \times
10^{-4}$~mag.
   
This allows us to convert from $\mathrm{rms}_{by}$ in magnitudes to
$\mathrm{rms_{white}}$ in relative flux.  Assuming a straightforward
proportionality relationship the conversion factor is $2.78$. This
conversion introduces a significant error in the overall conversion,
as we observed that the dependency of $\mathrm{rms_{white}}$ on
$R'_{\mathrm{HK}}$ (in the Sun) is slightly different in shape to that
of $\mathrm{rms}_{by}$ on $\left< R'_{\mathrm{HK}}\right>$ (in the
stellar sample), so a proportionality factor is inaccurate.  However,
until other stellar photometric time series that are sufficiently
regular to perform the fitting process described in Sect.\
\ref{sec:sbox} are available, it is the best we can do.  Finally one
must convert from $\mathrm{rms_{white}}$ to $A_1$. As expected, a
linear relationship between these quantities as computed for the Sun
is observed, yielding an overall conversion between
$\mathrm{rms_{by}}$ and $A_1$:
\begin{equation}
 \label{eq:a12rms}
      A_1 \times 10^5 = -0.24 + 0.66~\mathrm{rms}_{by}
\end{equation}
\noindent Eq.~(\ref{eq:rms2y}) thus becomes:
\begin{equation}
 \label{eq:a12y}
      A_1 \times 10^5 = 2.20 + 3.04~\left< R'_{\mathrm{HK}} \right>_5
        + 1.90~\left< R'_{\mathrm{HK}} \right>_5^2
\end{equation}

As space-based time-series photometric missions come online, we will
be able to calibrate the relations above with more and more
stars. Amongst these missions which have already provided useful data
or will start doing so very soon are MOST (Microvariability and
Oscillations of Stars, \citealt{wmk+03}), a small canadian mission
which saw first light on 20 July 2003, and the OMC (Optical Monitor
Camera, \citealt{gmj+99}) on board ESA's new $\gamma$-ray observatory
INTEGRAL. The possibility of using existing data from the star-tracker
camera of NASA's WIRE (Wide Field Infrared Explorer) is also under
investigation. This data has already been used to perform
asteroseismolgy on a number of bright stars \citep{cab+02} following
the failure of the main instrument shortly after launch. In the longer
term, the French/European collaboration COROT (COnvection, ROtation
and planetary Transits, \citealt{bag+03}), likely to be the first to
detect transits of (hot) rocky planets, will also provide a wealth of
information on stellar micro-variability.

These new data will allow us to calibrate the relations above with a
wider variety of stars. In particular, chromospheric activity and
rotational period measurements are available for a relatively large
number of bright late type stars
\citep{hbd+96,bss96,rls+98,hbd+00,tmj+02,psc+02}. If any of these are
observed by the missions listed above, yielding variability
measurements, they will be incorporated in the present model.

\subsection{Other parameters of the model}
\label{sec:other}

The previous sections were concerned with providing an estimate of one
of the model's parameters, $A_1$, given a star's age and colour.    
However, there are a total of 7 parameters to adjust. We have so
little information on the super- and meso-granulation component in
stars other than the Sun that we have chosen, for now, to leave it
unchanged in the simulations, using the solar values. 

\subsubsection{The third component: timescales of minutes or less}
  
The third (highest frequency) component observed in the Sun is a
superposition of variability on timescales of a few minutes, which is
thought to be related to granulation, and higher frequency effects
such as oscillations and photon noise. The distinction between these
effects is not resolved at the time sampling used for the present
study. There is very little information at the present time on
equivalent phenomena in other stars than the Sun (see below). Solar
values were therefore used in all simulated light curves for this
component. Given the low amplitude of this component, and the fact
that it corresponds to timescales significantly shorter than the
duration of planetary transits, it should not affect transit detection
significantly.

Granulation can be traced by studying asymmetries in line bisectors,
leading to a the possible exploration of this phenomenon across the HR
diagram (see for example \citealt{gn89}).  \citet{tcn+98}, who
modelled the granulation signal for the Sun, $\alpha$~Cen~A and
Procyon, obtained very similar power and velocity spectra for the
three stars, despite the fact that convection is much more intense on
Procyon. Although very preliminary, these results suggest that the
granulation power may change only slowly with stellar parameters, thus
supporting the use of the solar values in the present model.

\subsubsection{The second component: hours timescale}

Again, solar values were used for all simulated light curves for this
component. Work is underway to identify the types of surface
structures giving rise to the hours-timescale variability in the Sun
\citep{fsu00}, and the launch of the COROT mission will provide a
dataset ideally suited to improving our understanding of this type of
variability. Keeping the super- and/or meso-granulation component
identical to the solar case is certainly an oversimplification, and it
is the area where most effort will be focused in the future, as it is
highly relevant to transit detection, being on timescales similar to
transits.

\subsubsection{Timescale of the (first) active regions component}

Two parameters remain for the active regions component. We have kept
the slope of the power law, $C_1$, unchanged from the solar case.
Changing it slightly does not seem to affect the appearance of the
light curve significantly. More crucial is the timescale $B_1$.

\begin{figure}
  \centering \includegraphics[width=\linewidth]{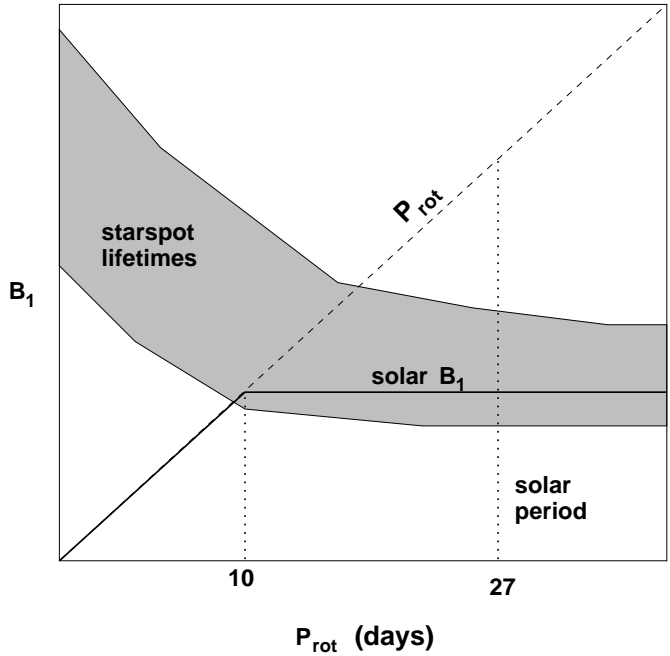}
  \caption{Schematic illustration of the expected dependence of
    the active regions component's characteristic timescale $B_1$ on
    rotational period. \label{fig:b1}}
\end{figure}

If the active regions component of micro-variability is the result of
the rotational modulation of active regions, we expect $B_1$ to be
directly related to the period.  However, the value obtained for the
Sun is $B_1 = 8.5 \times 10^6$~s, i.e.\ $9.84$ days, compared to a
rotational period of $\approx 26$~days. This suggests that the
timescale is not (or not exclusively) dominated by rotational
modulation of active regions, but by the emergence and disappearance
of the structures composing the active regions. Individual active
regions evolve on a timescale of weeks to months in the Sun
\citep{rls+98}, but spots and faculae evolve faster: observed sunspot
lifetimes range roughly between 10 days and 2 months \citep{hir02}.
We therefore deduce that sunspot evolution is the process which
determines $B_1$ in the Sun.

This reasoning can be generalised in the following way: the phenomenon
with the shortest timescale (in the relevant range) is expected to
determine $B_1$. In the Sun, this phenomenon is sunspot evolution. As
spot lifetimes are, if anything, longer in faster rotators
\citep{bcu+98,sfb99}, rotational modulation should take over below a
certain period. As a rough estimate we have placed the boundary
between the two regimes at $P_{\rm rot}=10$ days (see
Fig.~\ref{fig:b1}).

This completes, within the obvious limits of the assumptions used, the
requirements for the simulation of white light stellar light curves
with micro-variability, within the range of applicability of the
scaling laws used: $0.45 \leq B-V \leq 1.3$, $t \geq
t_{\mathrm{Hyades}}$, and the star must still be on the main sequence.
The stellar parameters required are age (or rotation period) and $B-V$
colour (or spectral type). It is also possible to supply $A_1$ and
$B_1$ directly. Once an artificial power spectrum is generated, phases
drawn at random from a uniform distribution are applied before
applying a reverse Fourier transform to return to the time domain.

Due to the use of randomly chosen phases, the shape of the variations
does not resemble the observed solar variability. It may be
possible to characterise the sequence of phases characteristic of a
given type of activity-related event, such as the crossing of the
stellar disk by a starspot, facula or active region. This information
could then conceivably be included, with appropriate scaling, in the
model, in order to make the shape of the variations more realistic.
However, how to do this in practice is not immediately obvious, and
will be investigated in the future. As the model stands, the simulated
light curves can be used to estimate quantities such as amplitude,
timescale, the distribution of residuals from a mean level, but no
conclusions should be drawn from the shape of individual variations.

\subsection{A few basic tests}
\label{sec:tests}

\begin{figure}
  \centering \includegraphics[width=\linewidth]{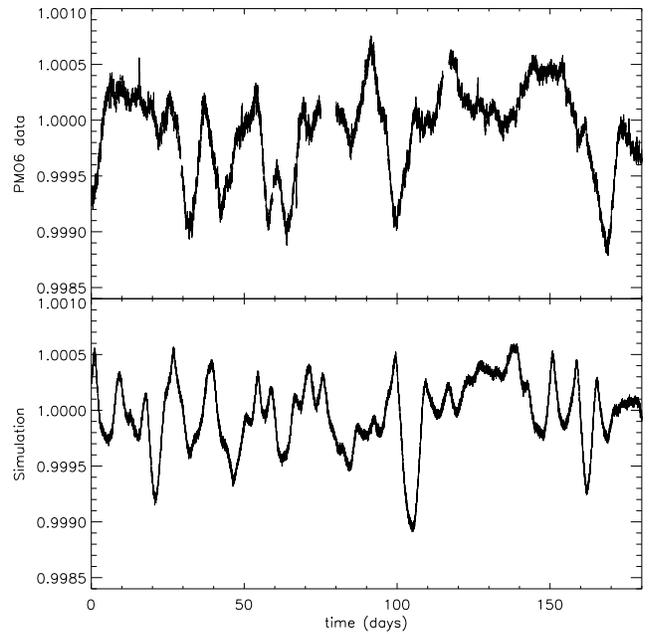}
  \caption{Comparison of a portion of PMO6 data starting 1300 days
    after the start of the full light curve (top panel) with a
    simulated light curve generated using the Sun's observed rotation
    period ($25.4$ days) and chromospheric activity index
    ($R'_{\mathrm{HK}} = -4.89$, \citealt{rls+98}).  Both light curves
    have 15 min sampling, last 180 days and are normalised to a mean
    flux of $1.0$. \label{fig:compsun}}
\end{figure}

A first check is to compare the predicted observables for the Sun to
the measured values. The measured period and $\log \left<
R'_{\mathrm{HK}} \right>$ are $25.4$~days and $-4.89$ \citep{rls+98},
in good agreement with the predicted values of $23.4$~days and
$-4.84$. The measured $\mathrm{rms}_{by}$ is $\simeq 5$ in units of
$10^{-4}$~mag. The value measured from a simulated light curve with
$1$~day sampling lasting $6$~months, when allowing for the conversion
between $\mathrm{rms_{white}}$ and $\mathrm{rms}_{by}$ is $6.11$. This
slight over-prediction is attributable to the Sun's slightly slow
rotation and low activity for its age and type, and to the fact that
it is slightly under-variable for its activity level (it falls
slightly below the fit on Fig.\ \ref{fig:rms2y}). Fig.\
\ref{fig:compsun} compares a portion of the PMO6 light curve taken
from a relatively high activity part of the solar cycle with a light
curve simulated using the observed rotation period and activity index
of the Sun. The amplitude and typical timescales of the variations are
well matched.

\begin{figure}
  \centering \includegraphics[width=\linewidth]{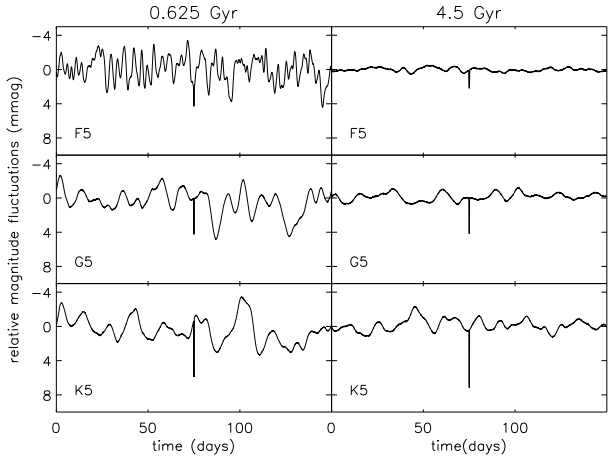}
  \caption{Examples of simulated light curves containing micro-variability
    for Hyades age (left column) and solar age (right column) stars
    with spectral types F$5$ (top row), G$5$ (middle row) and K$5$
    (bottom row). A single transit by a $0.5~R_{\mathrm{Jup}}$ planet
    has been added to each light curve 75 days after the start (a
    transit by an Earth-sized planet would be $\approx 25$ times
    smaller). The light curves have $1$~hr sampling and last 150 days.
    \label{fig:lcplot} }
\end{figure}

A set of six light curves have been simulated, corresponding to three
spectral types (F5, G5 and K5) and two ages (625 Myr and 4.5 Gyr).
Examination of the light curves and the various parameters computed
during the modelling process can reveal any immediate discrepancies.
The light curves are shown in Fig.\ \ref{fig:lcplot} and the
parameters in Table \ref{tab:lcpar}. They follow the expected trends,
variability decreasing with age and $B-V$ and increasing with
$P_{\mathrm{rot}}$, so that at a given age the least active star is
the G star, while the most active is the K star, despite is long
rotation period (due to the dependence of activity on colour).

\begin{table} 
  \caption[]{Parameters of the simulated light curves. \label{tab:lcpar}}
     $$ 
         \begin{array}{llllllll}
            \hline
            \noalign{\smallskip}
            \mathrm{Age} & \mathrm{SpT} & B-V & P_{\mathrm{rot}} & 
            \log \left( R'_{\mathrm{HK}} \right) & A_1 & 
            B_1 & \mathrm{rms_{white}} \\
            \mathrm{Gyr} & & & \mathrm{days} & & \times 10^5 & 
            \mathrm{days} & \times 10^4 \\
            \noalign{\smallskip}
            \hline
            \noalign{\smallskip}
            0.625 & \mathrm{F}5 & 0.44 &  2.9 & -4.64 & 19.27 & 2.89 & 22.0 \\
            0.625 & \mathrm{G}5 & 0.68 &  8.8 & -4.44 & 38.16 & 8.80 & 46.5 \\
            0.625 & \mathrm{K}5 & 1.15 & 12.6 & -4.42 & 41.40 & 9.84 & 69.5 \\
            4.5   & \mathrm{F}5 & 0.44 &  7.8 & -5.23 &  4.66 & 7.87 & 13.3 \\
            4.5   & \mathrm{G}5 & 0.68 & 23.7 & -4.80 & 11.95 & 9.84 & 13.9 \\
            4.5   & \mathrm{K}5 & 1.15 & 33.8 & -4.67 & 17.53 & 9.84 & 14.4 \\
            \noalign{\smallskip}
            \hline
         \end{array}
     $$ 
\end{table}


\section{Impact of micro-variability on transit detection from space}
\label{sec:simul}

A set of routines have been written to allow the rapid and automated
generation of light curves containing variability, some or no
transits, and photon noise, given the age and spectral type of
the star, if applicable the radius, period and mass of
the planet, and the expected photon count per sampling time. This
allows the systematic exploration of the detection limits in the
stellar and planetary parameter space for a given instrument.

For our study we have used the current configuration of the
\emph{Eddington} mission, as described by \citet{fav03}. The mission
is currently (June 2003) in its detailed definition phase, so that the
final payload configuration may evolve from the current baseline.
Currently, the \emph{Eddington} payload is constituted by a set of 4
identical wide field telescopes, with identical white light CCD cameras
at the focal plane. In practice, this is equivalent to a single
monolithic telescope with the same collecting area as the sum of the
collecting areas of the 4 individual telescopes. The total collecting
area of the baseline payload design is 0.764 m$^2$, and the field of
view has a diameter of 6.7~deg. The CCD chips used will be from E2V
and will have the standard ``broad band'' response. The mission is
scheduled for a launch in early 2008, and will perform its terrestrial
planet-finding program by observing a single region of the sky for
three years without interruptions.

\subsection{Which are the best target stars -- 
Impact on the choice of planet-finding field}
\label{sec:smany}

The aim here is to identify, for example, those stars which are likely
to be so variable that the detection of terrestrial planets orbiting
them by \emph{Eddington} and \emph{Kepler} will be seriously hindered,
or on the contrary where the transits are easily recovered even in the
presence of variability.  This will be used to optimise both the
choice target fields and the observing strategy, so that the range of
apparent magnitudes containing most of the best target stars is well
covered.

Light curves were therefore generated for a grid of star ages
($0.625$, $1.0$, $2.0$, $3.0$ \& $4.5$ Gyr) and types (F$5$, F$8$,
G$0$, G$2$, G$5$, G$8$, K$0$, K$2$ \& K$5$). Planetary transits were
added to the light curves for $1$ \& $3~R_{\oplus}$ and
$1~R_{\mathrm{Jup}}$ planets with periods of $30$~days, $6$~months,
$1$~year and $3$~years, resulting in $37$, $7$, $3$ and $1$ transit(s)
respectively. The light curves last $3$~years and have a sampling of
$1$~hr. Photon noise was added as suitable for the current
\emph{Eddington} baseline design.
Two apparent magnitudes, $V=13$ and $V=15$, were used.

Note that the expected sampling rate is in fact closer to 10 min than
1 hr. However, this first set of simulations was designed to rapidly
explore the stellar parameter space to identify regions of interest.
For this purpose, the light curves were generated using a longer
integration time, thereby keeping them to a manageable
size and maximising the contribution of stellar noise relative to
photon noise on a given data point -- the impact of photon noise
having already been investigated in a previous paper \citep{af02}.
Later simulations, concentrating on the habitable zones of the more
promising target stars, were made with 10 min sampling.

\begin{figure}
  \centering \includegraphics[width=\linewidth]{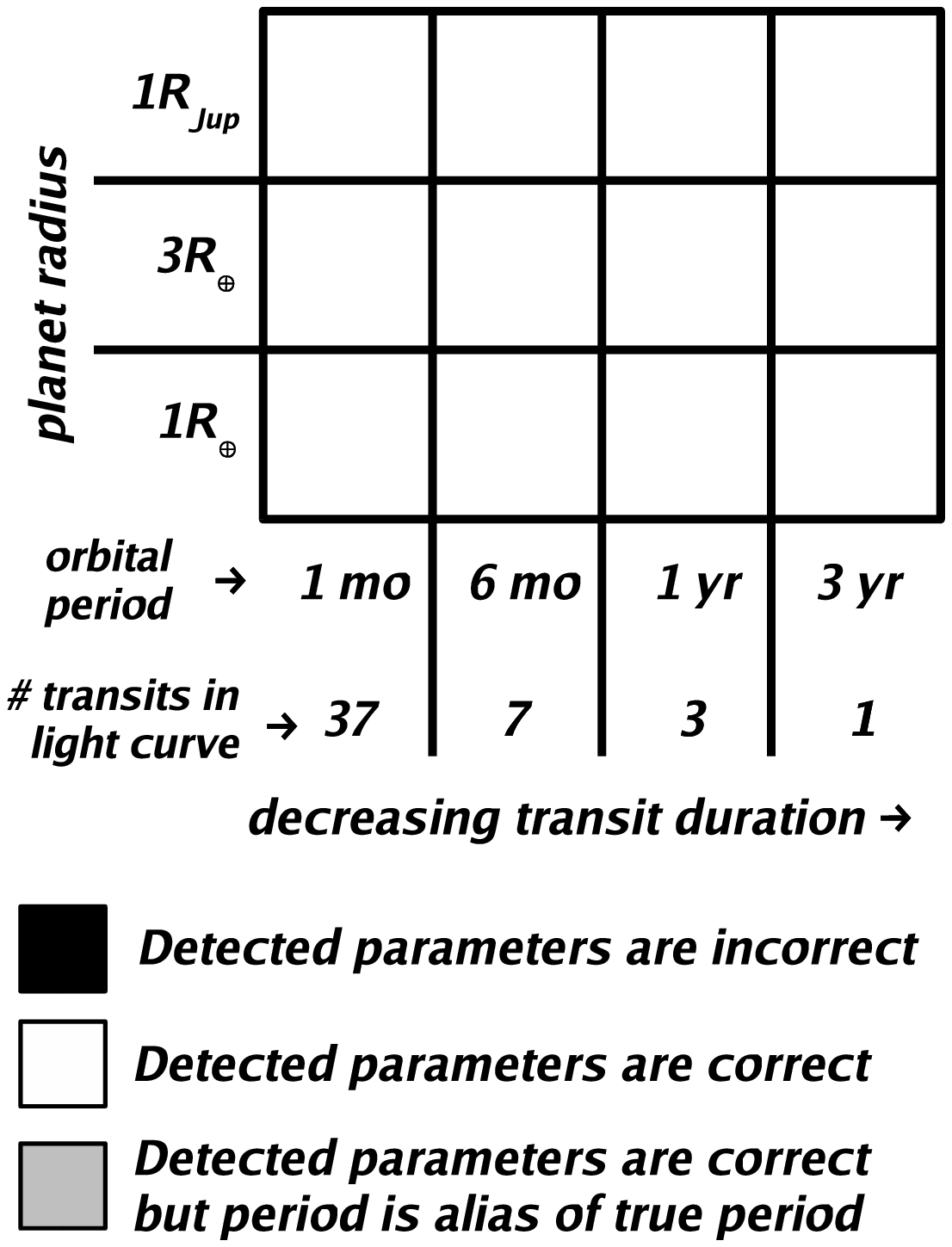}
  \caption{Notation used to present the results of the \emph{Eddington}
    simulations. \label{fig:not}}
\end{figure}

\begin{figure}
  \centering \includegraphics[width=\linewidth]{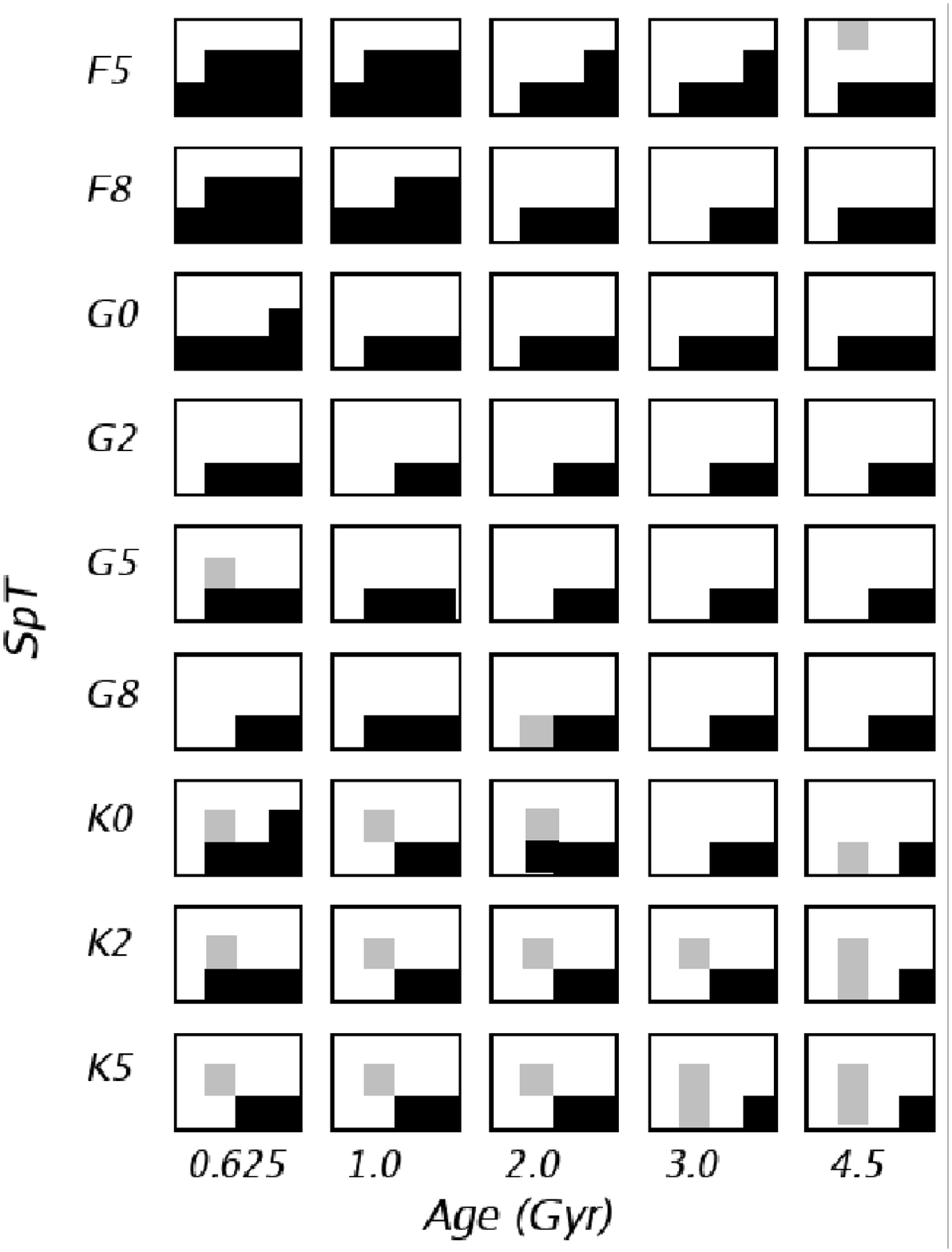}
  \caption{Results of the \emph{Eddington} simulations with $1$~hr sampling
    and $V=13$. The notation used is detailed in Fig.\ \ref{fig:not}.
    \label{fig:v13}}
\end{figure}

\begin{figure}
  \centering \includegraphics[width=\linewidth]{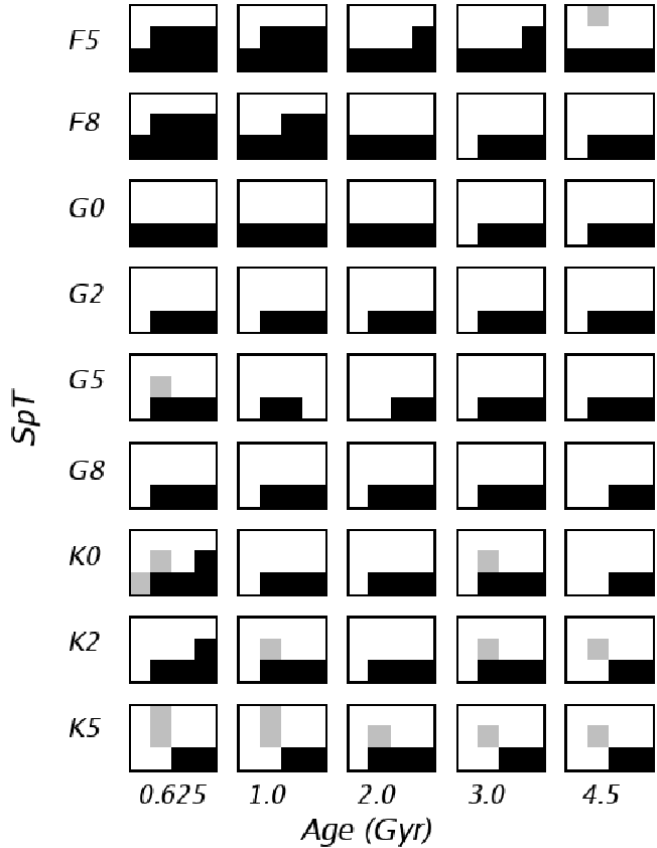}
  \caption{Results of the \emph{Eddington} simulations with $1$~hr sampling
    and $V=15$. The notation used is detailed in Fig.\ \ref{fig:not}.
    \label{fig:v15}}
\end{figure}

To estimate the detectability of the transits in each case, the light
curves were first pre-processed using an optimal Wiener-like filter
\citep{caf03}. This has the effect of whitening the noise and
increasing the signal to noise ratio of the transits. A transit
detection algorithm based on a maximum likelihood approach with a
simple box-shaped transit model \citep{ia03} was then applied to the
filtered light curves.  The results are shown in Figs.\ \ref{fig:v13}
\& \ref{fig:v15}, using the notation described in Fig.\ \ref{fig:not}.
For each case, we have indicated whether the best candidate transit(s)
(i.e.\ the set of trial parameters which gave rise to the minimum in
the detection statistic), corresponds to the true transit(s) inserted
in the light curve. Note that only one simulation was performed for
each star-planet combination, and the results should therefore be
taken as indicative rather than quantitative. The decision not to
perform full Monte Carlo simulations at this stage was taken because
both the variability model and the detection and filtering algorithms
are still undergoing frequent improvements.

The main conclusions to be drawn from these results are:
\begin{itemize}
\item The detection of transits by planets with $R_{\mathrm{pl}} \leq
  3~R_{\oplus}$ around stars younger than $2.0$~Gyr or earlier than
  G$0$ is significantly impaired, and they are not good targets for
  the exo-planet search of \emph{Eddington} or for \emph{Kepler}.
\item The small radius (hence increased transit depth) of K stars
  outweighs their relatively high variability levels, making them the
  best targets aside from effects not considered here, such as
  magnitude distribution and crowding. These will need to be assessed
  carefully. Whether this trend continues for M stars -- recalling
  that they were not included in the simulations because of the
  significant number of fast M rotators present at Hyades age -- is an
  open question \citep[see][]{dee03}. Another complicating factor for
  these late-type stars is the need to include the effects of micro-
  and nano-flaring, an issue under investigation.
\item Earth-sized planets are not detected correctly  around G stars
  with $3$ transits 
  only. This is only an indicative result, but it
  demonstrates the need to increase the transit signal-to-noise ratio
  for such systems, i.e.\ the square root of the number of in-transit
  points times the signal-to-noise ratio per data point. As photon
  noise is not the limiting noise factor at $V=13$, increasing the
  collecting area would not achieve this.  Instead, one needs to
  increase the number of in-transit points, while keeping the
  signal-to-noise ratio per data point constant.  This can be achieved
  through longer light curve duration. Finer time sampling would
  increase the number of in-transit points, but also the photon noise
  per data point. It is thus expected to improve the results for the
  bright, stellar noise dominated stars. The magnitude limit between
  the stellar noise and photon noise dominated regimes depends on the
  integration time, so that the optimum sampling rate is likely to be
  magnitude dependent.
\item At $V=15$ and with $1$~hr sampling, photon noise has become the
  dominant factor for most stars.
\end{itemize}
   
The third point in the list above could have important consequences
for the target field/star selection for \emph{Kepler}. The current
target field is centred on a galactic latitude of $\simeq 5^{\circ}$
in the region of Cygnus, and was chosen to maximise the number of
stars in the field while being sufficiently far from the plane of the
ecliptic to allow continuous monitoring \citep{bkd+97}. Due to
telemetry constraints, which limit the number of observed stars to
$\simeq 100\,000$, and to crowding, which can have a very large impact in such
high density regions, only stars with $V \leq 14$ are likely to be
monitored in this field. However, with its larger collecting area
(corresponding to a single aperture of $0.95$~m), Kepler's light
curves will be star noise rather than photon noise dominated down to
fainter magnitudes than in the case of \emph{Eddington}. Given that,
out of the spectral types tested here, K stars gave the best results,
and that these stars tend to be fainter than earlier types, it may be
desirable to choose a field at higher galactic latitude, combined with
a deeper magnitude limit. It would likely be difficult to extend the
magnitude limit while keeping the same target field due to increased
crowding.
   
The final choice of target fields for both \emph{Eddington} and
\emph{Kepler} will depend on many factors besides micro-variability
and the change of stellar radii with spectral type, which are the only
two effects taken into account in the present simulations. The
constraints imposed by limited telemetry budgets, as well as the
different apparent magnitude distributions of different stellar types,
and the problems due to crowding in high density regions, have already
been mentioned. It is only possible to characterise the spectral types
of the stars in the field through multi-colour photometry -- a
necessary task if efficient target star selection is to be achieved,
and one that would require very large amounts of telescope time if
attempted spectroscopically -- if patchy extinction is avoided.
Full-sky surveys based on photographic plates, such as the Digitized
Palomar Observatory Sky Survey \citep{djc+02}, can be used to detect
regions of patchy extinction.  Another important consideration is the
availability of ground-based facilities accessible to the scientific
community involved with each mission, for preparatory and follow-up
observations. As the target field will be far from the ecliptic plane,
it will be efficiently observable from either the southern or the northern
hemisphere, but not both.

\subsection{Minimum detectable transit in the presence of variability}
\label{sec:radius}

Another question of interest is whether \emph{Eddington} will really
probe the habitable zone of the stars it targets, if micro-variability
is taken into account. For this purpose, we have simulated light
curves with increased sampling ($10$~min) for three $4.5$~Gyr old
stars (types G$2$, K$0$ \& K$5$) orbited by planets with radii of
$0.8$, $1.0$, $1.5$ \& $2.0~R_{\oplus}$, with periods of $180$, $360$
\& $365$~days (corresponding to $7$, $4$ and $3$ transits in the light
curves respectively). The last two have very similar periods (i.e.\ 
virtually identical transit shapes and durations) but the extra
transit in the $360$~day case is due to the fact that the first
transit occurs soon after the start of the observations. This is
intended to test the effect of adding an extra year to the
planet-search phase, in order to detect more transits.

The boundary between the stellar noise dominated and photon noise
dominated regime, situated between $V=13$ and $V=15$ for 1 hr
integration, will be approximately $1$~mag brighter for $10$~min
integrations. In order to ensure that this second set of simulations
still covered, at least in part, the stellar noise dominated regime,
two values were used for the apparent magnitude of the stars: $V=12$
and $V=13$.
\begin{figure}
  \centering \includegraphics[width=\linewidth]{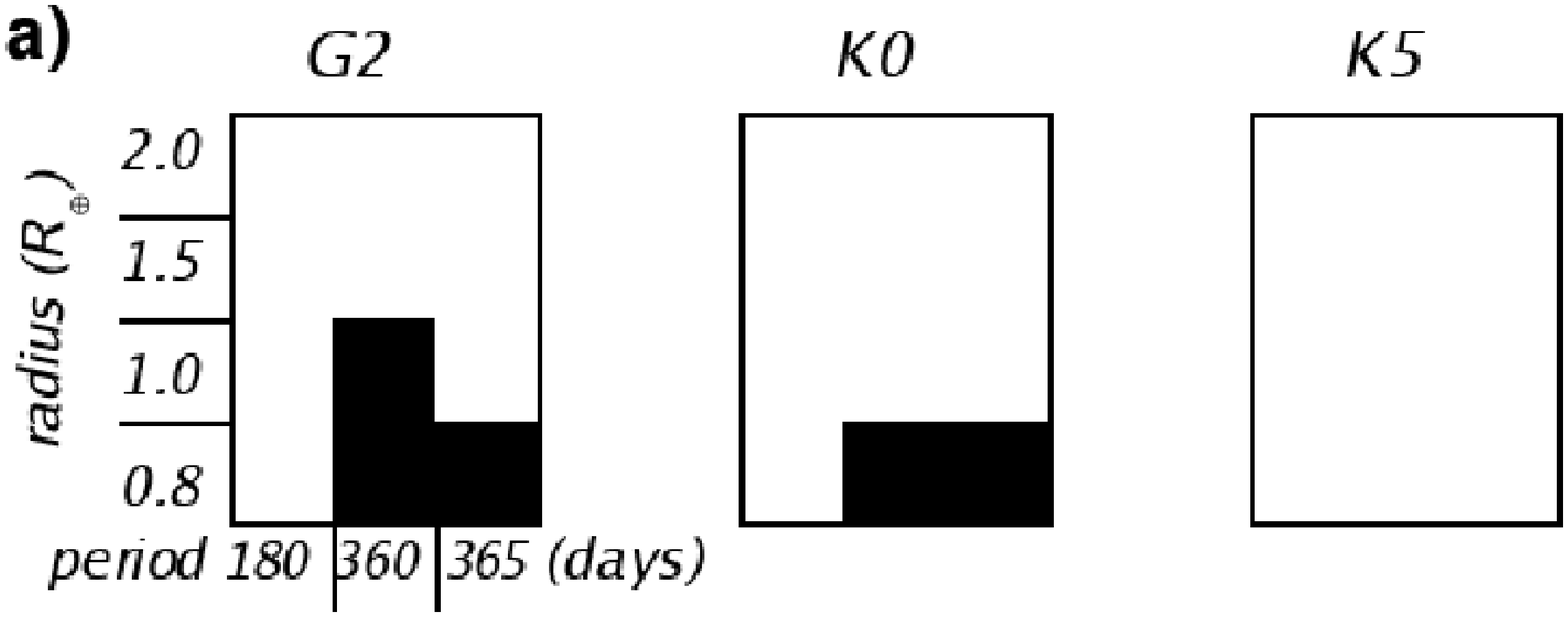}
  \centering \includegraphics[width=\linewidth]{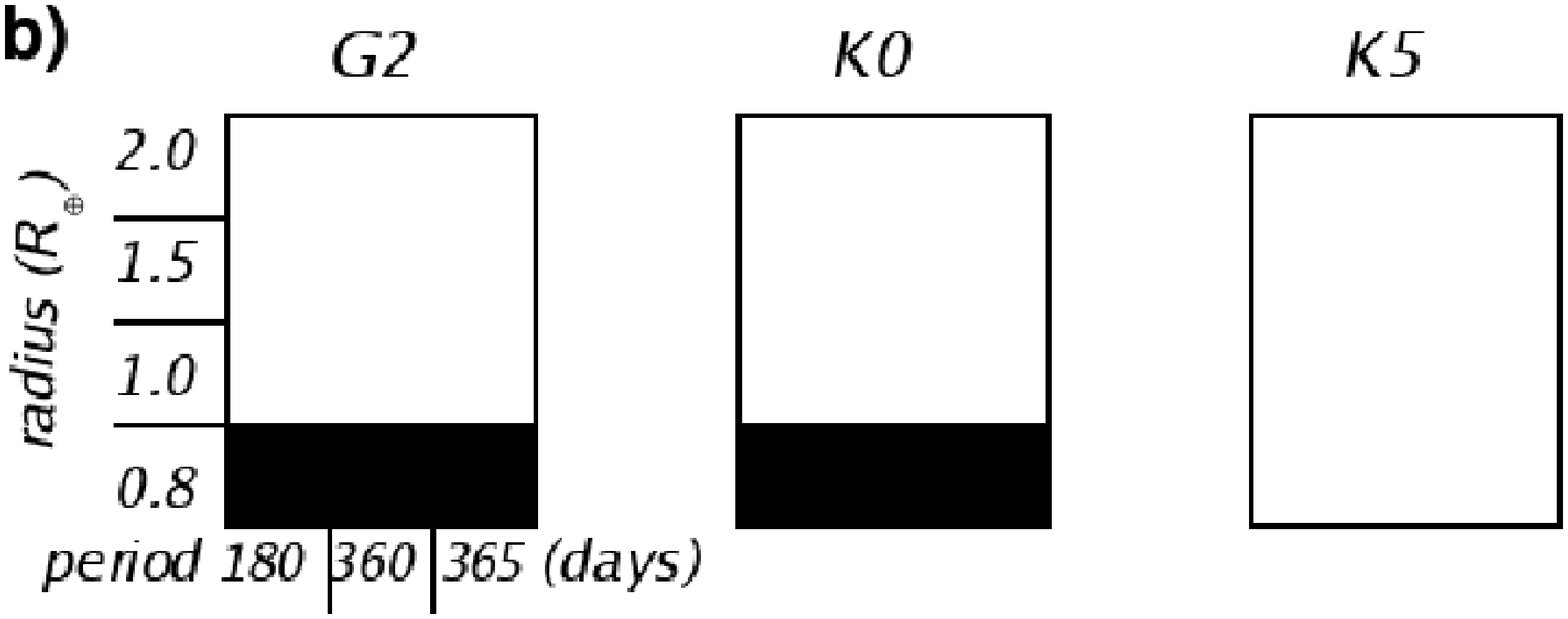}
  \caption{Results of the \emph{Eddington} simulations with 
    $10$~min sampling and apparent stellar magnitudes of $V=12$ {\bf
      (a)} and $V=13$ {\bf (b)}. \label{fig:rad}}
\end{figure}

The results are shown in Fig.\ \ref{fig:rad}. Generally speaking, they
follow the expected trends. While the $0.8~R_{\oplus}$ planet is
detected provided enough transits are present at $V=12$, it is not
detected around all but the smallest star at $V=13$, as a result of
the increased level of photon noise. The Earth-sized planet is only
marginally detectable with 3 or 4 transits around a G$2$ star, even at
$V=12$. The fact that it is detected with 4 transits but not with 3 at
$V=12$, and is detected in both cases at $V=13$, highlights the need
for full Monte Carlo simulations to obtain more reliable detectability
estimates. It should serve as a reminder that the present work only
aims to provide a global picture of the trends with star, planet and
observational parameters, rather than quantitative results. Closer
inspection of the light curve containing 4 Earth-sized transits around
a G$2$ star at $V=12$ showed that two of them were superposed on parts
of the light curve where the noise was consistently positive, which
impeded the detection. The rate of such coincidences can only be
estimated from multiple realisations of the same system. To summarise,
in the stellar noise dominated regime, the minimum reliably detectable
radii for G$2$, K$0$ and K$5$ stars are roughly $1.5$, $1.0$ and
$0.8~R_{\oplus}$ respectively with 3 or 4 transits, and
$0.8~R_{\oplus}$ in all cases with 7 transits.

By comparison, `habitable' planets are usually required to have radii
in the range $0.8~R_{\oplus} \leq R_{\mathrm{pl}} \leq 2.2~R_{\oplus}$.
This choice of radii, based on the rationale of \citet{bkd+97},
corresponds to a mass range of $0.5$ to $10~M_{\oplus}$. Above
$10~M_{\oplus}$, accretion of H and He starts and the planet will likely
become a gaseous giant. Below $0.5~M_{\oplus}$, the planet is not able
to retain a significant atmosphere, or to sustain plate tectonics on
biologically significant timescales. They must also lie in the
`habitable zone'.  This zone is usually defined to ensure the
existence of liquid water on the surface of the planet, the inner edge
being due to loss of water through photolysis and hydrogen escape, and
the outer edge to an increase in albedo which leads to a drop in
surface temperature \citep{kwr93}. \citet{fbb+02} compute
age-dependent habitable zone limits, requiring the surface temperature
to be in the range $0$ to $100~^{\circ}$C and taking into account the
change in luminosity of a star while on the main sequence. They give
the following orbital distance ranges for the habitable zones of
$4.5$~Gyr old stars: $0.5$ to $1$~A.U.\ ($0.8~M_{\odot}$, K$2$ star)
and $0.9$ to $1.3$~A.U.\ ($1.0~M_{\odot}$, G$2$ star). A similar but
simplified calculation (K.\ Horne, priv.\ comm.)  using a single
luminosity value, and converting to period using the mass range $0.5$
to $10~M_{\oplus}$, yields the periods corresponding to the centre of
the habitable zone as $1.2$, $0.6$ and $0.3$~years for G$2$, K$0$ and
K$5$ stars respectively.

The effect of the increased sampling rate is immediately visible.
While a `true Earth analogue' (Earth-sized planet orbiting a G$2$ star
with a period of $1$~year) may not be detected reliably, a good part
of the habitable zone of the G$2$ star and all of that of the K$0$ and
K$5$ stars are covered. This suggests that the primary goal of
discovering and characterising extra-solar planets in the habitable
zone will be achievable around intermediate age and old late-G and K
stars with the current design. To push back these limits,
modifications to the baseline design -- such as the possible inclusion
of colour information -- are under study at present. The fact that the
detectability of both $1.0$ and $0.8~R_{\oplus}$ planets increases
significantly with increased number of transits, suggests that it may
be desirable to increase the duration of the planet search stage, or
to return to the planet search field for confirmation after a break
(during which the asteroseismology programme, the other primary goal
of \emph{Eddington}, would be carried out).

\subsection{Detection threshold definition}
\label{sec:th}

An interesting consequence of introducing non-Gaussian noise in the
data is illustrated by these simulations. The fact that the detection
was correct does not imply it would have been considered significant.
Given a light curve in which there may or may not be one or more
transits, some kind of threshold must be used, in order to
automatically exclude candidate transits which are likely to be
spurious. Theoretically, in the presence of Gaussian noise only, a
global threshold can be used, i.e.\ a detected transit can be
considered reliable if its signal to noise exceeds a certain value,
whatever the star or planet parameters \citep{jcb02}. However, the use
of a single threshold (designed to keep the number of false alarms to
1 or less) in the present case leads to an alarmingly high rate of
missed detections. In particular, almost all correctly detected
transits by Earth-sized planets would be discarded because their
signal to noise ratio is too low. This is presumably due to the
non-Gaussian nature of the stellar noise.  However, it is interesting
to note that the detected duration is almost always the maximum trial
value when the detected event is spurious. This suggests that the
spurious events which mimic transits last, on average, longer than
transits, and might be a pointer to a method for excluding most of the
spurious events automatically.  A possible solution to the S/N
threshold problem might therefore be found by taking into account the
duration of the detected event when computing a detection threshold --
and applying a more stringent criterion to the longer duration events.
This possibility will be investigated in the near future.

For the purposes of the present simulations, the correct or incorrect
detection of the single realisation of a given star-planet system was
taken as indicative of the detectability of such a system by
\emph{Eddington}. To evaluate more precisely expected detection limits
once the planet finding field have been chosen, more detailed
simulations can be carried out.


\section{Conclusions}
\label{sec:concl}

A model to generate artificial light curves containing intrinsic
variability on timescales from hours to weeks for stars between mid-F
and late-K spectral type and older than $0.625$~Myr has been
presented. This model relies on the observed correlation between the
weeks timescale power contained in total solar irradiance variations
as measured by VIRGO/PMO6 and the Ca\,{\sc ii} K-line index of
chromospheric activity. The resulting light curves appear consistent
with currently available data on variability levels in clusters and
with solar data. Further testing and fine-tuning requires high
sampling, long duration space-based stellar time-series photometry and
will be carried out as such data become available.

The simulated light curves can be used to estimate the impact of
micro-variability on exo-planet search missions such as
\emph{Eddington} and \emph{Kepler}. The results of simulations
including stellar micro-variability, photon noise as expected for
\emph{Eddington} and transits have been presented. After optimal
filtering, the results of transit searches on these light curves
suggest that stellar micro-variability, combined with the change in
stellar radius with spectral type, will make the detection of
terrestrial planets around F-type stars very difficult. On the other
hand, K-stars appear to be promising candidates despite their high
variability level, due to their small radius. 

At $V=12$ and with $10$~min sampling, the smallest detectable
planetary radii for $4.5$~Gyr old G$2$, K$0$ and K$5$ stars, given a
total of 3 or 4 transits in the light curves, were found to be $1.5$,
$1.0$ and $0.8~R_{\oplus}$ respectively. This result was obtained in
stellar noise rather than photon noise dominated light curves, and
therefore also applies to lower apparent magnitudes or larger
collecting areas. The magnitude limit beyond which photon noise would
start to dominate, thereby increasing the minimum detectable radii for
a given star and observing configuration, depends on the collecting
area and sampling time, but the effects of increased photon noise are
detected at $V=13$ for $10$~min sampling time and \emph{Eddington}'s
collecting area.

All the simulated light curves produced so far, as well as the
routines used to generate them, have been made available to the
exo-planet community through the web page: \linebreak
\verb\www.ast.cam.ac.uk/~suz/simlc\. Light curves with specific
parameters can be generated on request. The inclusion of this model in
the simulating tools under development for both missions is under
study.

The next step is the extension of the model presented here to include
colour information, to allow a detailed assessment of the pros and
cons of including broadband filters on one or more of
\emph{Eddington}'s four telescopes. This will be the subject of an
upcoming paper. At the same time, the performance of a range of
filtering and transit detection algorithms, which have recently been
compared in simulations including white noise only \citep{tin03}, will
be compared in the presence of micro-variability.


\begin{acknowledgements}
  We wish to thank G.\ Micela for her help regarding the calibration
  of the scaling laws in Sect.\ \ref{sec:model} and S.\ Hodgkin for
  his careful reading of the manuscript and helpful comments. The PMO6
  data was made available by T.\ Appourchaux and the SoHO/VIRGO
  team. SoHO is a mission of international collaboration between ESA
  and NASA. The BBSO \ Ca\,{\sc ii} K-line index records were provided
  by T.\ Henry, S.\ Keil and the BBSO staff. Much of the development
  of the variability filter and transit detection algorithm used in
  Sect.\ \ref{sec:simul} are due to S.\ Carpano and M.\ Irwin
  respectively.  S.\ Aigrain was supported by PPARC Studentship number
  PPA/S/S/2001/03183 and a studentship from the Isaac Newton Trust.
\end{acknowledgements}

\bibliographystyle{aa} 
\bibliography{aigrain}

\end{document}